\documentclass[aps,prb,reprint,groupedaddress]{revtex4-2}

\usepackage{graphicx}% Include figure files
\usepackage{booktabs}

\begin{document}
\title{Modified Kondorsky Domain Reversal in Microstructured Phase-Separated Manganites} 

\author{Monique Kubovsky}
\email{mkubovsky@ufl.edu}
\affiliation{Department of Physics, University of Florida, Gainesville, Florida 32611, USA}
\author{Dylan Tagrin}
\affiliation{Department of Physics, University of Florida, Gainesville, Florida 32611, USA}

\author{Amlan Biswas}
\email{amlan@phys.ufl.edu}
\affiliation{Department of Physics, University of Florida, Gainesville, Florida 32611, USA}

\date{\today}

\begin{abstract}
The hole-doped manganite (La$_{1-y}$Pr$_{y}$)$_{0.67}$Ca$_{0.33}$MnO$_3$ (LPCMO) shows electronic phase separation between ferromagnetic metallic (FMM) and anti-ferromagnetic charge-ordered insulating (AFM-COI) regions. In this study, (La$_{0.5}$Pr$_{0.5}$)$_{0.67}$Ca$_{0.33}$MnO$_{3}$ (LP5CMO) microstructures were fabricated using photolithography on thin films grown on (110) NdGaO$_3$ (NGO) substrates. We investigated the domain reversal mechanism of these microstructures through magnetotransport measurements. Our results demonstrate that, while bulk (unpatterned) films follow the standard Kondorsky model for domain reversal, the microstructures obey a modified Kondorsky model. This difference indicates that local magnetic fields from reversed domains significantly influence the coercive field in confined geometries. Although we did not observe a strong electric field effect, this study establishes that magnetotransport measurements are a feasible method for probing the competition between shape and magnetocrystalline anisotropy in manganite microstructures, which could provide an alternative path for controlling magnetic domains at low current densities.
\end{abstract}

\pacs{}
%Magnetic domains, Dielectrophoresis, Magnetic anisotropy, Magnetoelectric effects  

\maketitle 

\section{Introduction}

The ability to control the coupling between magnetism and charge transport through electric field effects could reveal novel phases not accessible through conventional materials-preparation techniques~\cite{rmp24}. Such electric-field-induced states may also enable new types of data-storage applications for magnetic materials~\cite{Par08,Zha24,Gro20}. Strongly correlated oxides (SCO) and two-dimensional magnetic van der Waals materials (2D magnets) are both promising candidates for tuning magnetic behavior with applied electric fields due to their rich phase diagrams with multiple competing phases~\cite{rmp24}. 2D magnets also offer confined dimensionality, which can unveil otherwise hidden functionalities~\cite{natcom25}. However, advanced fabrication and lithography developed for strongly correlated oxides, such as manganites, can produce high-quality, dimensionally confined samples that are more stable than 2D magnets, enabling us to combine competing phases, confinement, and mature processing to explore electric-field-driven behavior in SCOs~\cite{prm25}. Within this class of materials, manganites such as (La$_{y}$Pr$_{1-y}$)$_{1-x}$Ca$_{x}$MnO$_{3}$ (LPCMO), exhibit phase coexistence between the ferromagnetic metallic (FMM) and anti-ferromagnetic charge-ordered insulating (AFM-COI) regions~\cite{Jee11, Ahn04, Ueh99, Mur10}. Phase separation, together with substrate-induced epitaxial strain, provides alternative pathways to tune magnetic properties, such as the coercive field and magnetic anisotropy, through electric field effects like dielectrophoresis~\cite{Sha20, Don07, Don10}. 

There are two main sources of magnetic anisotropy (MA), namely shape anisotropy (SA) and magnetocrystalline anisotropy (MCA). SA is a classical effect, determined by factors such as reduced dimensionality, while MCA is due to the quantum effect of spin-orbit coupling~\cite{Cul08, hong2016, blank2009}. Previous work on electric field tuning of magnetic properties focused primarily on microscopic quantum effects with less emphasis on classical effects at the mesoscopic (micrometer) scale~\cite{rmp24}. 

In LPCMO thin films grown on (110) NdGaO$_3$ substrates, the anisotropic in-plane strain {\em and} phase separation leads to in-plane magnetic anisotropy with the easy axis along the higher tensile strain $(1\bar{1}0)$ direction~\cite{blank2009, Jee11, Sin12a}. The anisotropic strain also leads to anisotropic percolation under an applied electric field with preferential (faster) percolation along the magnetic easy axis~\cite{Jee13}. This behavior is due to dielectrophoresis of isolated, single-domain FMM regions which are slightly elongated along the magnetic easy axis ~\cite{Jee13,Jee11,Don10,Mur10}. This suggests that the SA of the FMM regions in LPCMO plays a role in determining the overall MA and reveals an alternate route for electric field tuning of MA. To investigate this possibility, we need to tune the SA while keeping the MCA fixed, which necessitates the fabrication of LPCMO microstructures similar in length scale to that of the FMM regions in the phase separated state~\cite{Zha02,Sin12b}. However, the magnetic properties of such microstructures cannot be measured directly using magnetization measurements because of the small sample volumes. Therefore, alternative methods are needed to measure the magnetic properties, such as coercive fields, of these microstructures. 

\begin{figure*}[t]
\includegraphics[width=\textwidth] {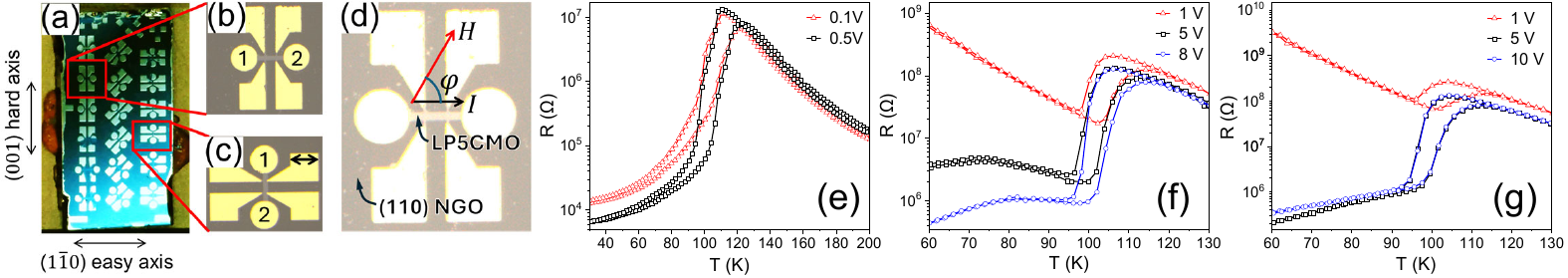}
\caption{Optical microscope image showing the (a) samples in the chip carrier, (b) easy axis sample, and (c) hard axis sample. The double-sided arrow in (c) shows the 100 $\mu$m width of a gold contact. (d) The microstructure in (b) digitally brightened to show that the LP5CMO microstructure on the NGO substrate. The width of the LP5CMO bars (light gray) is 20 $\mu$m. Resistance vs. temperature behavior of (e) the unpatterned LP5CMO thin film, (f) the LP5CMO easy axis microstructure, and (g) the LP5CMO hard axis microstructure. The hysteresis between the warming (higher transition temperature) and cooling (lower transition temperature) cycles is observed for all the samples due to phase coexistence.}
\label{Microstructures}
\end{figure*}

In this work, we used magnetotransport measurements to track the domain reversal mechanisms of the FMM regions in microstructures of (La$_{0.5}$Pr$_{0.5}$)$_{0.67}$Ca$_{0.33}$MnO$_{3}$ (LP5CMO) thin films grown on (110) NdGaO$_3$ (NGO) substrates. We used photolithography to fabricate LP5CMO wires of dimensions 20 $\times$ 100 $\mu$m$^2$, with the length aligned along either the magnetic easy or hard axis. We also measured the effect of in-plane electric fields on the FMM domain dynamics using these device structures. The sensitivity of domain wall scattering (DWS) and tunneling effects to the orientation of the FMM domains allowed us to estimate the coercive field ($H_C$) of the microstructures from resistance vs.\ magnetic field ($R(H)$) measurements, instead of direct magnetization measurements, which were precluded due to the small volume of the microstructures~\cite{Sun_2018,Frydman_2005,Wu_2020,Yu_2021}. The measured coercive fields were similar to those of bulk (unpatterned) LP5CMO thin films~\cite{Sin23}. We varied the angle ($\theta$) between the easy axis and the applied magnetic field to generate $H_C$ vs.\ $\theta$ plots that followed a modified Kondorsky model suggesting that domain reversal occurs due to a different type of nucleation and growth process compared to unpatterned manganite thin films~\cite{Urz22,Kon40,Jee11,Sin23,Wan03}. Although we did not observe a significant electric field effect on the domain dynamics in LP5CMO, likely due to a high fraction of the FMM phase, our results show that it is feasible to measure the electric field effect on the competition between SA and MCA in manganite microstructures using magnetotransport measurements.

\section{Experimental methods}

LP5CMO thin films of nominal thickness 30-nm were grown on NGO substrates using pulsed laser deposition (PLD)~\cite{Jee11}. A two-step photolithography process was then used to fabricate microstrucures of the LP5CMO thin films. In the first step, 800 nm of AZ1512 photoresist was spin-coated onto the thin films. Hall-bar patterns were transferred to the films by exposure to UV light through a photomask using a Karl Suss MA6 contact aligner. The samples were then developed, wet-etched using an HCl and KI solution, and the photoresist was removed to define the LP5CMO microstructures. In the second step, titanium and gold electrical contacts were deposited on the microstructures using lift-off photolithography. Photoresist was applied again, and the sample was re-exposed to UV light through a photomask with the pattern of contacts. The sample was then baked in ammonia to reverse the photoresist. After developing the sample, 10 \AA\ of titanium and 100 \AA\ of gold was sputtered to form the electrical contacts using a KJL Multi-Source RF/DC Sputter System. After the photoresist was removed, the samples were mounted and gold wire bonded to a chip carrier. The finished sample is shown in Fig.  \ref{Microstructures}(a). Microstructure morphology and thickness were characterized by optical and atomic force microscopy (Digital Instruments Nanoscope III). The sample was then loaded into a custom-built cryostat which accommodates the chip-carrier and enables rotation of the sample current with respect to a fixed magnetic field. The cryostat was inserted into an American Magnetics cryostat with a 9 T magnet for the magnetotransport measurements. For the resistance vs.\ temperature ($R(T)$) measurements, the samples were cooled down to 30 K and a voltage was then applied across the samples (using indium contacts for the unpatterned thin film and across terminals labeled 1 and 2 in Fig. \ref{Microstructures}(b) and (c) for the microstructures). $R(T)$ data were taken during both the warming and cooling runs. For $R(H)$ measurements at a specific temperature and electric field, the patterned sample was cooled in zero magnetic field from 125 K to the target temperature with a voltage $V_{\mathrm{cool}}$ applied across the terminals labeled 1 and 2 in Fig. \ref{Microstructures}(b) and (c). $R(H)$ measurements were then taken at various angles $\varphi$ between the sample voltage/current and the magnetic field. $\varphi$ was changed while at the target temperature and in zero magnetic field before the next magnetic field sweep.

\section{Results and discussion}

\subsection{Resistance vs. Temperature Behavior of LP5CMO Microstructures}

Fig. \ref{Microstructures}(a) shows the optical microscope image of the fabricated LP5CMO microstructures, with the easy axis sample (for voltage/current along the easy axis) in Fig. \ref{Microstructures}(b) and the hard axis sample (for voltage/current along the hard axis) in Fig. \ref{Microstructures}(c). The gold-titanium (Au-Ti) electrical contacts appear yellow in these images. The double-sided arrow in Fig. \ref{Microstructures}(c) serves as a scale bar as it shows the 100 $\mu$m width of a gold contact. Fig. \ref{Microstructures}(d) shows the same microstructure as Fig. \ref{Microstructures}(b) but has been digitally brightened to show that the LP5CMO microstructure has the dimensions of 20 $\times$ 100 $\mu$m$^2$ between the circular contacts. These circular Au-Ti contacts, labeled 1 and 2 in Fig. \ref{Microstructures}(b) and (c), were used to apply the voltage along the length of the sample and measure its longitudinal resistance. The geometry of the contacts ensures that the current is confined along either the easy or hard axis. The square contacts were deposited to measure the transverse Hall voltage but were not used for the present set of experiments. Fig. \ref{Microstructures}(d) also defines $\varphi$ as the angle between the magnetic field $H$ and current $I$. Hence, the angle $\theta$ between $H$ and the easy axis direction is the same as $\varphi$ for the easy axis sample and is $90^{\circ}-\varphi$ for the hard axis sample. We will use $\theta$ to define the orientation of $H$ for the remainder of this paper.

$R(T)$ measurements of both the bulk (unpatterned) thin film and the microstructures were taken to determine the insulator-to-metal transition temperatures $T_{IM}$, which we defined as the temperature at which the resistance peaks during the cooling run. The warming cycle has a higher transition temperature, resulting in the hysteresis observed for all the samples due to phase coexistence. Fig. \ref{Microstructures} (e) shows that the bulk film has a $T_{IM}$ around 110 K, and Fig. \ref{Microstructures} (f) and (g) show that the microstructures have $T_{IM}$s of around 106 K. The similarity of the $T_{IM}$s confirms that the photolithography preserved the essential properties of the manganite. However, the slight difference between these results demonstrates that current confinement along either the easy or hard axis has an effect on the transport properties of the material due to the percolative nature of the transition at $T_{IM}$~\cite{Ueh99}. We can further see effects of current confinement in the behaviors of the easy and hard axis samples below $T_{IM}$. The transition from insulator to metal causes the current to rise about two orders of magnitude, which leads to local heating and an increase in resistance with cooling below 98 K. Dielectrophoresis is more effective in aligning FMM regions along the easy axis, so the heating effect is more apparent for the easy axis sample, especially for the two higher voltages~\cite{Jee13}. In the 1 V measurements, the insulating behavior remains at low temperature due weaker dielectrophoretic effect at lower electric fields~\cite{Jee13, Sha20}.

\begin{figure}
\includegraphics[width=0.5\textwidth] {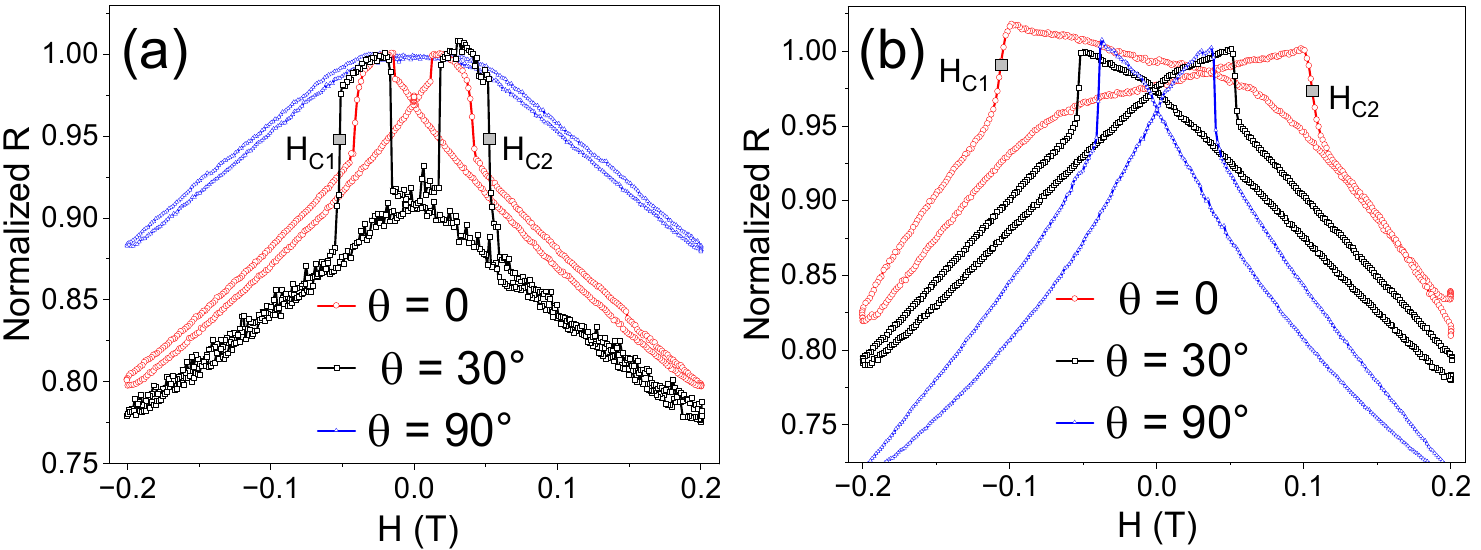}
\caption{Normalized resistance versus applied magnetic field data taken at 98 K with $V_{\mathrm{cool}}= 5$ V used to determine $H_C$ for both (a) the easy and (b) hard axis devices. $H_{C1}$ and $H_{C2}$ are marked in the figures.
\label{RawRvH}
}
\end{figure}

\subsection{Magnetoresistance and Coercive Fields of the Microstructures}

The $R(T)$ measurements were followed by $R(H)$ measurements to determine how microstructuring and applied voltages (and hence, electric fields) affect the $H_C$ of the material. The $R(H)$ measurements were taken at $T= 98$K and 94 K for different values of $V_{\mathrm{cool}}$. These two temperatures were chosen based on the $R(T)$ data which showed that the voltage (electric field) effect was strongest at these temperatures. The peak resistance observed in $R(H)$ measurements occurs at $H_C$ due to spin-dependent tunneling between the FMM regions separated by the AFM-COI phase, which acts as a tunneling barrier~\cite{Sun_2018}. The highest resistance state is reached when the magnetic moments of the FMM regions are randomly oriented, maximizing spin-dependent scattering for the tunneling electrons~\cite{Yu_2021}. Furthermore, even in a single FMM region, a resistance maximum in $R(H)$ is linked to the reorganization of the magnetic domain structures (which includes domain wall motion/scattering) as the material reverses magnetization~\cite{Sun_2018}. This coupling between a material's magnetic state and its electrical resistance allows the estimation of the $H_C$ of the LP5CMO microstructures by locating the field of the resistance maximum in the $R(H)$ data. We defined the negative and positive coercive fields ($H_{C1}$ and $H_{C2}$, respectively) as the field values at which the resistance drops are the sharpest, as shown in Fig. \ref{RawRvH}. The $H_C$ was then defined as the average $(H_{C2}- H_{C1})/2$ with an error bar of $(H_{C2}+ H_{C1})/2$ ($H_{C1}$ is negative). 

The slight elongation of the FMM regions along the easy axis results in their preferential percolation in this direction~\cite{Jee13,Don10}. Consequently, for transport measurements with $V_{\mathrm{cool}}$ applied along the easy axis, the FMM regions are separated by insulating domain walls which are remnants of the AFM-COI phase and are sufficiently narrow to permit quantum tunneling~\cite{Sin09}. This gives rise to a magnetotransport behavior consistent with tunneling magnetoresistance, characterized by an initial resistance increase at lower fields followed by sharp drops at $H_{C1}$ and $H_{C2}$ (Fig. \ref{RawRvH}(a)) ~\cite{Sin09b}. The selection of these higher magnetic fields ($H_{C1}$ and $H_{C2}$) is justified by their proximity to the coercive field values observed in unpatterned LP5CMO films~\cite{Sin23,Jee11}. In contrast, for transport with $V_{\mathrm{cool}}$ applied along the hard axis, the intervening AFM-COI regions are too wide to allow tunneling. The magnetotransport is therefore dominated by domain wall scattering, which manifests as resistance drops at $H_{C1}$ and $H_{C2}$ without the preceding resistance increase (Fig. \ref{RawRvH}(b)). 

In addition to the resistance drops at $H_{C1}$ and $H_{C2}$, the $R(H)$ plots reveal other features relevant to determining the magnetotransport mechanisms in these microstructures. Fig. \ref{RawRvH} shows that the negative background magnetoresistance is approximately linear for different configurations of the current/voltage and magnetic field directions relative to the magnetic easy axis, except when both the current/voltage and the magnetic field are along the hard axis, where it becomes quadratic (Fig. \ref{RawRvH} (b)). Negative linear magnetoresistance is observed in magnetic films, such as epitaxial Fe$_3$O$_4$, due to spin-polarized transport across atomically sharp antiferromagnetically coupled antiphase boundaries within the film~\cite{Eer02,Ram06}. Negative quadratic magnetoresistance is also observed under similar conditions when the magnetic field is applied along the magnetic hard axis and is less than the anisotropy field~\cite{Eer02,Zha15,Suc22}. In this low-field regime, field-induced tilting of antiferromagnetic spins leads to increased conductivity (negative magnetoresistance) that is proportional to the square of the applied magnetic field~\cite{Eer02,Suc22}. These observations are consistent with our data on LP5CMO microstructures, with the added condition that both the current and magnetic field must be along the hard axis for quadratic magnetoresistance to occur. This distinction is also likely due to anisotropic percolation, which leads to wider antiferromagnetic boundaries separating the FMM regions along the hard axis~\cite{Jee13,Don10}.

\begin{figure}
\includegraphics[width=0.5\textwidth] {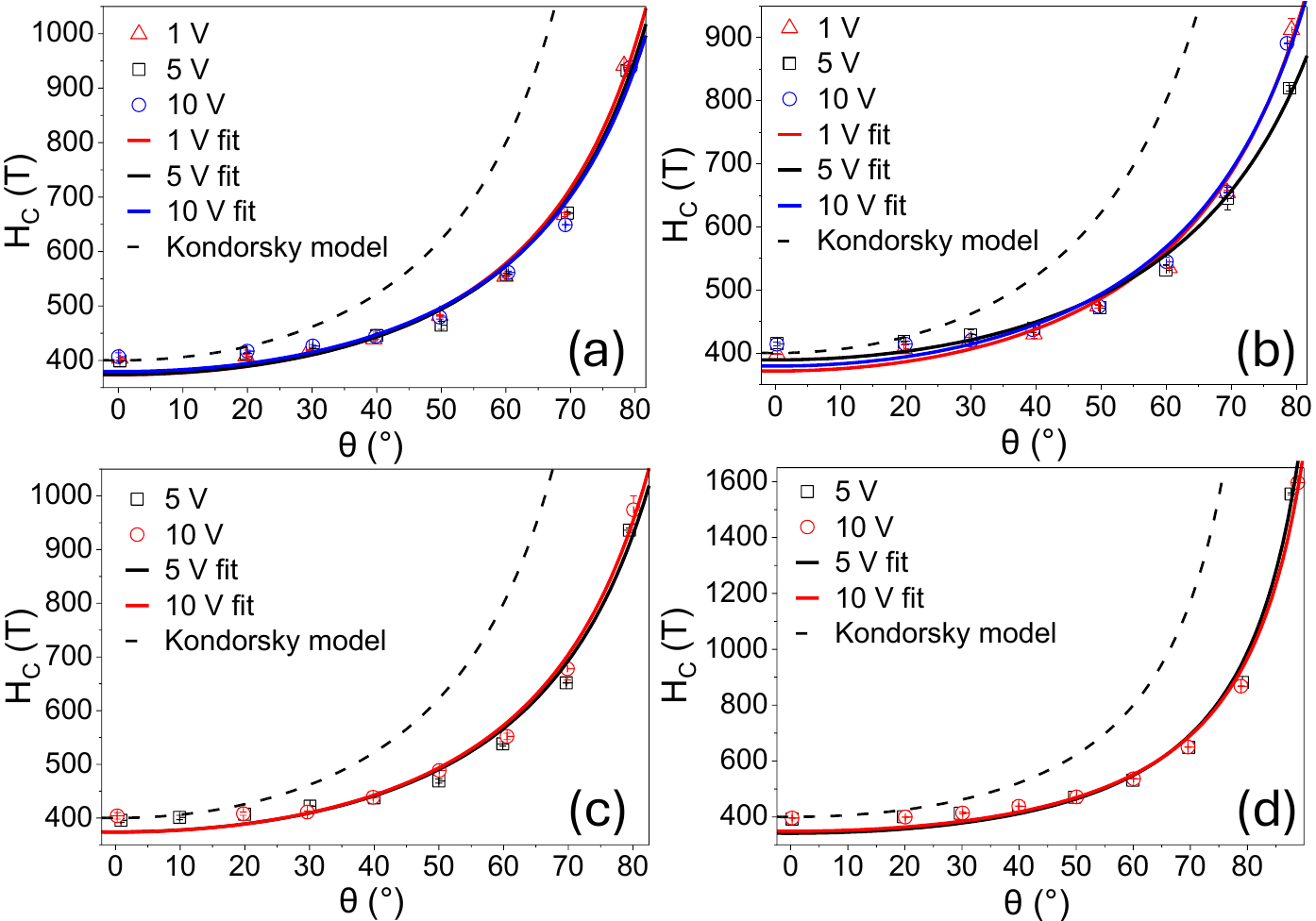}
\caption{$H_C$ versus $\theta$ for the easy axis sample at (a) 94K and (b) 98K and for the hard axis sample at (c) 94K and (d) 98K, and the corresponding modified Kondorsky model fits. The Kondorsky model is also shown as a dashed line for comparison.}
\label{modkon}
\end{figure}

\begin{table} [t]
    \begin{center}
    \caption{Parameters for the modified Kondorsky model used to fit the $H_C(\theta)$ behavior of LP5CMO microstructures.}
    \begin{tabular}{cccc@{\hskip 0.2in}c@{\hskip 0.1in}cc}
       \toprule
           $I$ Direction & $T$ (K) & $V_{\mathrm {cool}}$ (V) & $H_{C0}$ (G) & $k$ & $M_s$ \\
       \midrule
       Easy &  94 &  1 &      370 $\pm$ 10      & 4.3 $\pm$ 0.4  & 6.9 $\pm$ 0.8\\
       Easy &  94 &  5 &      370 $\pm$ 10      & 4.1 $\pm$ 0.4  & 7.3 $\pm$ 0.8\\
       Easy &  94 &  10 &     380 $\pm$ 11      & 3.8 $\pm$ 0.3  & 8.0 $\pm$ 1.0\\
       Easy &  98 &  1 &      370 $\pm$ 10       & 3.8 $\pm$ 0.3  & 7.8 $\pm$ 0.9\\
       Easy &  98 &  5 &      389 $\pm$ 8       & 2.9 $\pm$ 0.2  & 11 $\pm$ 0.9\\
       Easy &  98 &  10 &     380 $\pm$ 10      & 3.6 $\pm$ 0.3  & 8.0 $\pm$ 1.0\\
       Hard &  94 &  5 &      370 $\pm$ 10       & 3.8 $\pm$ 0.3  & 7.8 $\pm$ 0.9\\
       Hard &  94 &  10 &     374 $\pm$ 9       & 4.1 $\pm$ 0.3  & 7.3 $\pm$ 0.7\\
       Hard &  98 &  5 &      340 $\pm$ 14      & 5.3 $\pm$ 0.4  & 5.1 $\pm$ 0.6\\
       Hard &  98 &  10 &     350 $\pm$ 12      & 4.8 $\pm$ 0.3  & 5.7 $\pm$ 0.7\\
       \bottomrule
    \end{tabular}
    \label{tab:fitting_parameters}
\end{center}
\end{table}

\subsection{Variation of Coercive Field with $\theta$: A Modified Kondorsky Model}

To further probe the influence of the microstructuring and cooling voltage $V_{\mathrm{cool}}$ on the magnetic properties of LP5CMO, we measured the coercive field as a function of the angle $\theta$ between the applied magnetic field and the easy axis ($H_C(\theta)$). This angular-dependent analysis allows us to directly assess how $V_{\mathrm{cool}}$ affects the magnetic anisotropy and the underlying domain reversal mechanisms within the microstructures. Bulk (unpatterned) manganite thin films, including LP5CMO, follow the Kondorsky model of domain reversal via nucleation and propagation, as represented by equation \ref{Kondorsky model} ~\cite{Cul08,Kon40,Jee11,Sin23,Wan03}:
\begin{equation}
         H_C(\theta)=\frac{H_{C0}}{\cos{\theta}}
      \label{Kondorsky model}
\end{equation}
where $H_{C0}$ is the minimum coercive field observed when the magnetic field is along the easy axis. In contrast to the behavior of the unpatterned manganite thin films, Fig. \ref{modkon} shows that the microstructures of LP5CMO do not follow the Kondorsky model. Instead, the variation of $H_C$ with $\theta$ follows a modified Kondorsky model ~\cite{Hys13,Urz22}. The simplest explanation for the modified Kondorsky model is the local magnetic field created by the magnetization of reversed domains in the material, which reduces the global coercive field required for domain reversal~\cite{Urz22}. According to this model, $H_C$ is further modified due to the role defects play in domain nucleation and growth~\cite{Sko14,Schu91}. In an unpatterned film, the effects of local magnetic fields and defects appear to be negligible, and the domain nucleation and propagation follow the standard Kondorsky model. However, in the microstructures, our data suggest that the local magnetism is strong enough to lower the coercive field. To obtain a quantitative picture of the domain reversal mechanism in the microstructures, we fit the $H_C(\theta)$ curves to the modified Kondorsky model represented by equation \ref{MFK} ~\cite{Urz22, Aco24}:
\begin{equation}
         H_C(\theta)=\frac{H_{C0}}{\cos{\theta^*}}
      \label{MFK}
\end{equation}
where
\begin{equation}
         \theta^*= \arcsin{\left (\frac{k \sin{\theta}}{\sqrt{k^2 +2k \cos{\theta} +1 }}\right )}
      \label{theta*}
\end{equation}
and
\begin{equation}
         k=\frac{H_{C0}}{4\pi M_s}
      \label{k parameter}
\end{equation}
In equations~\ref{theta*} and ~\ref{k parameter}, $k$ is a fitting parameter dependent on $H_{C0}$ and a parameter $M_s$, which is related to the local magnetic field caused by the reversed magnetic domains~\cite{Urz22, Aco24}. Table \ref{tab:fitting_parameters} shows the fitting parameters $k$ and $H_{C0}$ obtained by fitting our data to the modified Kondorsky model. The $k$ and $H_{C0}$ values were used to calculate $M_s$. We compared these $M_s$ values to the fully spin-polarized saturation magnetization for LP5CMO with hole-doping of 33\% at low temperatures. The theoretical value is 3.67 $\mu_B$/Mn, or 592 emu/cm$^3$, which is close to the experimental value of about 580 emu/cm$^3$~\cite{Sin23}. The $M_s$ values in Table \ref{tab:fitting_parameters} are much lower than the saturation magnetization of LP5CMO, which is consistent with the fact that $M_s$ is the local magnetization due to only the reversed magnetic domains. The $k$ values show negligible or inconsistent variation with the magnitude and direction of $V_{\mathrm{cool}}$, suggesting that there is no significant effect of electric fields on the MA of LP5CMO microstructures. This insensitivity of the MA to electric fields is possibly due to the high ($\sim50\%$) fraction of the FMM phase in LP5CMO just below $T_{IM}$, which does not allow setting up the non-uniform electric fields required for dielectrophoresis~\cite{Sin23, Jee13, Sha20}. 

\section{Conclusions}

We have successfully fabricated microstructures of the electronically phase-separated manganite (La$_{0.5}$Pr$_{0.5}$)$_{0.67}$Ca$_{0.33}$MnO$_{3}$ (LP5CMO) using photolithography and demonstrated that these microstructures preserve the essential properties of the bulk material while exhibiting distinct effects due to current confinement along specific crystallographic axes. The key finding is that while bulk LP5CMO films follow the standard Kondorsky model for magnetic domain reversal, the microstructures instead follow a modified Kondorsky model, indicating that local magnetic fields from reversed domains significantly influence the coercive field and domain reversal mechanisms in confined geometries. Although in this study we did not observe significant electric field effects on domain dynamics in LP5CMO (likely due to a high fraction of the ferromagnetic metallic phase), the results establish the feasibility of using magnetotransport measurements to investigate electric field effects on the competition between shape anisotropy and magnetocrystalline anisotropy in manganite microstructures. Future studies should be performed on microstructures of manganites with a lower FMM fraction such as (La$_{0.4}$Pr$_{0.6}$)$_{0.67}$Ca$_{0.33}$MnO$_{3}$.

\begin{acknowledgments}
	This work was supported by the University of Florida's University Scholars Program. 
\end{acknowledgments}

\end{document}